\documentstyle[eqsecnum,floats,preprint,aps]{revtex}
\tighten

\begin{document}

\newcommand{\vp}{\varphi}
\newcommand{\CP}{${\cal CP}$}
\newcommand{\C}{${\cal C}$}
\newcommand{\Parity}{${\cal P}$}
\newcommand{\B}{${\cal B}$}
\newcommand{\be}{\begin{equation}}
\newcommand{\ee}{\end{equation}}
\newcommand{\bea}{\begin{eqnarray}}
\newcommand{\eea}{\end{eqnarray}}
\newcommand{\PSbox}[3]{\mbox{\rule{0in}{#3}\includegraphics{#1}\hspace{#2}}}

\newcommand{\bref}[1]{(\ref{#1})}
\newcommand{\Lag}{{\cal L}}
\newcommand{\th}{\theta}
\newcommand{\thb}{\bar{\theta}}
\newcommand{\al}{\alpha}
\newcommand{\ald}{{\dot{\alpha}}}
\newcommand{\sm}{{\sigma^\mu}}
\newcommand{\sigb}[1]{\bar{\sigma}^{{#1}}}
\newcommand{\sgth}{\sigma^\vp}
\newcommand{\sgr}{\sigma^r}
\newcommand{\sgz}{\sigma^z}
\newcommand{\la}{\lambda}
\newcommand{\lab}{\bar{\lambda}}
\newcommand{\psb}{\bar{\psi}}
\newcommand{\phb}{\bar{\phi}}
\newcommand{\xib}{\bar{\xi}}
\newcommand{\chb}{\bar{\chi}}
\newcommand{\dmu}{{\partial_\mu}}
\newcommand{\dth}{{\partial_\vp}}
\newcommand{\dr}{{\partial_r}}
\newcommand{\eth}[1]{e^{{#1}\vp}}
\newcommand{\spinU}{\mbox{\scriptsize 
		$\left(\begin{array}{c} 1 \\ 0 \end{array} \right)$}}
\newcommand{\spinD}{\left(\begin{array}{c} 0 \\ 1 \end{array} \right)}
\newcommand{\mat}[4]{\left(\begin{array}{cc} 
			{#1} & {#2} \\ {#3} & {#4} \end{array}\right)}

\title{$N=1$ Supersymmetric Cosmic Strings}

\author{Stephen C. Davis$^{1}$\footnote[1]{S.C.Davis@damtp.cam.ac.uk}, 
Anne-Christine Davis$^{1}$\footnote[2]{A.C.Davis@damtp.cam.ac.uk} and
Mark Trodden$^{2}$\footnote[3]{trodden@ctpa04.mit.edu.~~ Also, Visiting 
Scientist, Brown University, Providence RI. 02912.}}

\address{~\\$^1$Department of Applied Mathematics and Theoretical Physics \\
University of Cambridge, CB3 9EW, UK.}

\address{~\\$^2$Center for Theoretical Physics \\ 
Laboratory for Nuclear Science and \\
Department of Physics \\
Massachusetts Institute of Technology \\
Cambridge, Massachusetts 02139, USA.}

\maketitle

\begin{abstract}

We investigate the microphysics of supersymmetric cosmic strings. 
In particular we focus on the vortices
admitted by $N=1$ supersymmetric abelian Higgs models. 
We find the vortex solutions and demonstrate that the two simplest 
supersymmetric cosmic string models admit fermionic superconductivity.
Further, by using supersymmetry transformations, we show how to solve for
the fermion zero modes giving rise to string superconductivity in
terms of the background string fields.

\end{abstract}

\setcounter{page}{0}
\thispagestyle{empty}

\vfill

\noindent DAMTP-96-107\hfill 

\noindent MIT-CTP-2612

\hfill Typeset in REV\TeX

\eject

\vfill

\eject

\baselineskip 24pt plus 2pt minus 2pt

\section{Introduction}
In the past twenty years it has become clear that topological defects in 
quantum field theories may play an essential role in the evolution of the
early universe. One important effect of these topological solitons 
is gravitational. The evolution of a network of cosmic
strings produced at a grand unified (GUT) phase transition provides a
possible origin for the seed density perturbations which became the large 
scale structure of the universe observed today\cite{structure}.

However, important early universe physics may also arise from the 
microphysics of topological defects. If a network of defects 
is produced just prior to
the electroweak phase transition, their interactions with the fields of the
standard electroweak theory form the basis for an electroweak baryogenesis 
scenario which is insensitive to the details of the phase 
transition\cite{DMEWBG}. 
Further, the spectrum of radiation from strings produced during a 
Peccei-Quinn\cite{PQ} symmetry breaking provides important bounds on the 
allowed values of any axion mass\cite{B&S}.
Finally, cosmology in the presence of topological defects is qualitatively
altered if the strings carry superconducting currents, as first suggested by
Witten\cite{Witten}. In particular, if a network of cosmic strings becomes 
superconducting, then the possibility of producing massive stable remnants 
(vortons) allows one to constrain the underlying particle 
physics theory by cosmological considerations (for a recent analysis 
see~\cite{vortons}).

In this paper, we investigate the microphysics of cosmic string solutions 
admitted by supersymmetric (SUSY) field theories. This is important for at 
least two
reasons. First, SUSY field theories include many popular 
candidate theories of physics above the electroweak scale. Second, the recent 
successes of duality in SUSY 
Yang-Mills theories may mean that the physics of nonperturbative solutions
such as topological solitons may be easier to understand than in
non-supersymmetric theories. As in early studies 
of non-SUSY defects\cite{Kibble}, we work in the context of the simplest 
models and in
particular with versions of the abelian Higgs model obeying the supersymmetry
algebra with one SUSY generator ($N=1$). We
demonstrate that the particle content and interactions dictated by 
SUSY naturally give rise to cosmic string superconductivity in these models. 
Further, by using SUSY transformations, we are able to 
find solutions for the fermion zero modes responsible for 
superconductivity in terms of the background string fields. 
Ours is not the first analysis of superconducting SUSY cosmic
strings. However, whereas earlier analyses\cite{Morris} have focussed on the 
complicated structure of the supersymmetrized $U(1)\times U(1)$ Witten 
model\cite{Witten}, 
here we demonstrate the presence of superconductivity in even the simplest
SUSY cosmic string theories. A special case of the solutions 
discussed in this paper has been obtained in a similar model by other 
authors\cite{Garriga} using different techniques.

The structure of the paper is as follows. In the next section we present the
$N=1$ SUSY abelian Higgs models. Such simple SUSY models are well-known in
particle physics (for example see Ref.~\cite{Fayet I}). However, we
believe the cosmological relevance of the solutions we explore here to be 
new. In order to make contact
with both the supersymmetry and cosmology literature, we employ both the 
superfield and component formalisms and repeat
a number of well-established facts
and conventions for the sake of clarity.
Spontaneous symmetry breaking (SSB)
in these models can be implemented in two distinct ways, leading to different
theories with different particle content. We call these distinct
models theory F and theory D respectively to refer to the origin of the
SSB term in the Higgs potential. 
In section three we focus on theory F. We demonstrate how
the cosmic string solution can be constructed in the bosonic sector and
derive the equations of motion for the fermionic zero modes. We then 
employ SUSY transformations to solve these equations
in terms of the background string fields. In section four we
repeat the analysis for theory D. The type of symmetry breaking in 
theory D is peculiar to theories with an abelian gauge group and we 
therefore expect theory F 
to be more representative of models with nonabelian gauge groups such as
grand unified theories. In section five we
check our results for the special case discussed in Ref.\cite{Garriga}. In 
fact, for theory D, the solutions are already of this special form.
Finally, in section six, we comment on the possible implications of our 
findings.

\section{Supersymmetric Abelian Higgs Models and SSB}
Let us begin by defining our conventions. Throughout this paper we use the
Minkowski metric with signature $-2$, the antisymmetric 2-tensor
$\epsilon_{21} = \epsilon^{12} = 1$, 
$\epsilon_{12} = \epsilon^{21} = -1$ and the Dirac gamma matrices in the
representation
\be
\gamma_\mu = \mat{0}{\sm}{\sigb{\mu}}{0} \ ,
\ee
with $\sm = (-1,\sigma^i)$, $\sigb{\mu} = (-1,-\sigma^i)$ and where
$\sigma^i$ are the Pauli matrices.

We consider supersymmetric versions of the spontaneously broken gauged
$U(1)$ abelian Higgs model. In superfield notation, such a theory consists of
a vector superfield $V$ and $m$ chiral superfields $\Phi_i$, ($i=1\ldots m$),
with $U(1)$ charges $q_i$. In Wess-Zumino gauge these may be 
expressed in component notation as

\begin{equation}
V(x,\th,\thb) = -(\th\sm\thb)A_\mu(x) + i\th^2\thb\lab(x) 
		- i\thb^2\th\la(x) + \frac{1}{2}\th^2\thb^2 D(x) \ ,
\label{vectorDef}
\end{equation}

\begin{equation}
\Phi_i(x,\th,\thb) = \phi_i(y) + \sqrt{2}\th\psi_i(y) + \th^2 F_i(y) \ ,
\label{chiralDef}
\end{equation}
where $y^\mu = x^\mu + i\th\sm\thb$.
Here, $\phi_i$ are complex scalar fields and $A_\mu$ is a vector field. These
correspond to the familiar bosonic fields of the abelian Higgs model.
The fermions $\psi_{i \alpha}$, $\lab_{\alpha}$ and
$\la_{\alpha}$ are Weyl spinors and the complex bosonic fields, $F_i$, and 
real bosonic field, $D$, are auxiliary fields. Finally, $\th$ and $\thb$ are
anticommuting superspace coordinates. In the component formulation of the 
theory one eliminates $F_i$ and $D$ via their equations of motion and
performs a Grassmann integration over $\th$ and $\thb$.
Now define

\bea 
D_\al & = & \frac{\partial}{\partial\th^\al} +
i\sigma^\mu_{\al \ald} \thb^\ald \dmu \ , \nonumber  \\
\bar{D}_\ald & = & -\frac{\partial}{\partial\thb^\ald} -
i\th^\al \sigma^\mu_{\al \ald} \dmu \ , \nonumber \\
W_\al & = & -\frac{1}{4}\bar{D}^2 D_\al V \ , 
\eea
where $D_\al$ and $\bar{D}_\ald$ are the supersymmetric covariant derivatives
and $W_\al$ is the field strength chiral superfield.
Then the superspace Lagrangian density for the theory is given by

\begin{equation}
{\tilde \Lag} = \frac{1}{4}
\left( W^\al W_\al|_{\th^2} + \bar{W}_\ald \bar{W}^\ald|_{\thb^2} \right) 
+ \left( \bar{\Phi}_i e^{g q_i V} \Phi_i \right)|_{\th^2\thb^2}
+ W(\Phi_i)|_{\th^2} +\bar{W}(\bar{\Phi}_i)|_{\thb^2} + \kappa D \ .
\label{susyLag}
\end{equation}
In this expression $W$ is the superpotential, a holomorphic function of the 
chiral superfields (i.e. a function of $\Phi_i$ only and not $\bar{\Phi}_i$) 
and $W|_{\th^2}$ indicates the $\th^2$ component of $W$.
The term linear in $D$ is known as the Fayet-Iliopoulos
term \cite{Fayet II}. Such a term can only be present
in a $U(1)$ theory, since it is not invariant under more general gauge
transformations. 

For a renormalizable theory, the most general superpotential is

\be
W(\Phi_i) = a_i \Phi_i + \frac{1}{2}b_{ij} \Phi_i\Phi_j 
			+ \frac{1}{3}c_{ijk} \Phi_i\Phi_j\Phi_k \ ,
\ee
with the constants $b_{ij}$, $c_{ijk}$ symmetric in  their indices.
This can be written in component form as

\be
W(\phi_i, \psi_j, F_k)
  = a_i F_i + b_{ij}\left(F_i\phi_j - \frac{1}{2}\psi_i\psi_j\right)
	+ c_{ijk} \left(F_i\phi_j\phi_k - \psi_i\psi_j\phi_k \right) 
\ee
and the Lagrangian \bref{susyLag} can then be expanded in Wess-Zumino gauge
in terms of its component fields using (\ref{chiralDef},\ref{vectorDef}).
The equations of motion for the auxiliary fields are

\begin{equation}
F^\ast_i + a_i + b_{ij}\phi_j + c_{ijk}\phi_j\phi_k = 0
\ee
and 
\be
D + \kappa + \frac{g}{2} q_i \phb_i \phi_i = 0 \ .
\label{deqn}
\end{equation}
Using these to eliminate $F_i$ and $D$ we obtain the Lagrangian density 
in component form as

\begin{equation}
\Lag = \Lag_B + \Lag_F + \Lag_Y - U \ ,
\label{nsusyLag}
\end{equation}
with
\begin{eqnarray}
\Lag_B &=& (D^{i\ast}_\mu \phb_i) (D^{i\mu} \phi_i)
		- \frac{1}{4} F^{\mu\nu}F_{\mu\nu} \ , \\
\Lag_F &=& -i\psi_i \sm D^{i\ast}_\mu \psb_i - i\la_i \sm \dmu \lab_i \ , \\
\Lag_Y &=& \frac{ig}{\sqrt{2}} q_i \phb_i \psi_i \la 
 	- (\frac{1}{2}b_{ij} + c_{ijk}\phi_k) \psi_i \psi_j + (\mbox{c.c.})
\ , \\
   U   &=& |F_i|^2 + \frac{1}{2}D^2  \nonumber \\
       &=& |a_i + b_{ij}\phi_j + c_{ijk}\phi_j\phi_k|^2 
+\frac{1}{2}(\kappa + \frac{g}{2} q_i \phb_i \phi_i)^2 \ ,
\label{Ueqn}
\end{eqnarray}
where $D^i_\mu = \dmu + \frac{1}{2}ig q_i A_\mu$ and 
$F_{\mu\nu} = \dmu A_\nu - \partial_\nu A_\mu$. 

Now consider spontaneous symmetry breaking in these theories.
Each term in the superpotential must be gauge invariant.
This implies that $a_i \neq 0$
only if $q_i =0$, $b_{ij} \neq 0$ only if $q_i + q_j =0$, and 
$c_{ijk} \neq 0$ only if $q_i + q_j + q_k=0$. 
The situation is a little more complicated than in non-SUSY theories, since
anomaly cancellation in SUSY theories implies the existence of more than one 
chiral superfield (and hence Higgs field). In order to break the gauge
symmetry, one may either
induce SSB through an appropriate choice 
of superpotential, or, in the case of the $U(1)$ gauge
group, one may rely on a non-zero Fayet-Iliopoulos term.

We shall refer to the theory with superpotential SSB (and, for simplicity, 
zero Fayet-Iliopoulos term) as theory F and
the theory with SSB due to a non-zero Fayet-Iliopoulos term as theory D. 
Since the
implementation of SSB in theory F can be repeated for more general gauge 
groups, we
expect that this theory will be more representative of general defect-forming
theories than theory D for which the mechanism of SSB is specific to the 
$U(1)$ gauge group.

\section{Theory F: Vanishing Fayet-Iliopoulos Term}
The simplest model with vanishing Fayet-Iliopoulos term 
($\kappa=0$) and 
spontaneously broken gauge symmetry contains three chiral superfields.
It is not possible to construct such a model with fewer superfields which 
does not
either leave the gauge symmetry unbroken or possess a gauge anomaly.
The fields are two charged fields $\Phi_\pm$, with respective $U(1)$ charges 
$q_\pm = \pm 1$, and a neutral field, $\Phi_0$. A suitable superpotential is 
then

\begin{equation}
W(\Phi_i) = \mu \Phi_0 (\Phi_+ \Phi_- - \eta^2) \label{susyW} \ ,
\end{equation}
with $\eta$ and $\mu$ real.
The potential $U$ is minimised when $F_i=0$ and $D=0$. This occurs when
$\phi_0=0$, $\phi_+ \phi_- = \eta^2$, and $|\phi_+|^2 = |\phi_-|^2$.
Thus we may write $\phi_\pm = \eta e^{\pm i\al}$, where $\alpha$ is some 
function. We shall now seek the Nielsen-Olesen\cite{NO} solution 
corresponding to an infinite straight cosmic string.  
We proceed in the same manner as for
non-supersymmetric theories. Consider only the bosonic fields (i.e. set the 
fermions to zero) and in cylindrical polar coordinates $(r,\vp, z)$ write 
 
\begin{eqnarray}
\phi_0 & = & 0 \ , \\
\phi_+ & = & \phi_-^\ast = \eta e^{in\vp}f(r) \ , \\
A_\mu & = & -\frac{2}{g} n \frac{a(r)}{r}\delta_\mu^\vp \ , \\
F_\pm & = & D = 0 \ , \\
F_0 & = & \mu \eta^2 (1 - f(r)^2) \ ,
\label{StringSol}
\end{eqnarray}
so that the $z$-axis is the axis of symmetry of the defect. The profile 
functions, $f(r)$ and $a(r)$, obey 

\begin{equation}
f''+\frac{f'}{r} - n^2\frac{(1-a)^2}{r^2} = \mu^2 \eta^2 (f^2 -1)f \ ,
\label{fEqn}
\end{equation}

\begin{equation}
a''-\frac{a'}{r} = -g^2 \eta^2(1-a)f^2 \ ,
\label{aEqn}
\end{equation}
with boundary conditions 

\bea
f(0)=a(0)=0 \ , \nonumber \\ 
\lim_{r\rightarrow \infty}f(r)=\lim_{r\rightarrow\infty}a(r)=1 \ .\nonumber
\eea
Note here, in passing, an interesting aspect of topological defects in
SUSY theories. The ground state of the theory is supersymmetric,
but spontaneously breaks the gauge symmetry while in the core of the defect the
gauge symmetry is restored but, since $|F_i|^2 \neq 0$ in the core, 
SUSY is spontaneously broken there. 

We have constructed a cosmic string solution in the bosonic sector of the 
theory. Now consider the fermionic sector.
With the choice of superpotential \bref{susyW} the component form of the
Yukawa couplings becomes

\begin{equation}
\Lag_Y = i\frac{g}{\sqrt{2}}
	 \left(\phb_+ \psi_+ - \phb_- \psi_-\right) \la
		 - \mu \left(\phi_0 \psi_+ \psi_- + \phi_+ \psi_0 \psi_- 
		+ \phi_- \psi_0 \psi_+ \right) + (\mbox{c.c.})
\end{equation}

As with a non-supersymmetric theory, non-trivial zero energy fermion
solutions can exist around the string. Consider the fermionic ansatz 

\be
\psi_i = \spinU \psi_i(r,\vp) \ ,
\ee
\be
\la = \spinU \la(r,\vp) \ .
\ee
If we can find solutions for the $\psi_i(r,\vp)$ and $\la(r,\vp)$ then, 
following Witten, we know that solutions of the form

\be
\Psi_i=\psi_i(r,\vp)e^{\chi(z + t)} \ , \
 {\tilde \Lambda}=\la(r,\vp)e^{\chi(z + t)} \ ,
\label{witteq}
\ee
with $\chi$ some function, represent left moving superconducting 
currents flowing along the string at the speed of light. Thus, the
problem of finding the zero modes is reduced to solving for the 
$\psi_i(r,\vp)$ and $\la(r,\vp)$.

The fermion equations of motion derived from \bref{nsusyLag} are four
coupled equations given by

\be
\eth{-i}\left(\dr -\frac{i}{r}\dth \right)\lab - \frac{g}{\sqrt{2}} \eta
		f \left(\eth{in}\psi_- - \eth{-in} \psi_+\right) = 0 \ ,
\label{fermeq1}
\ee
\be
\eth{-i}\left(\dr -\frac{i}{r}\dth \right)\psb_0 + i \mu \eta 
		f \left(\eth{in}\psi_- + \eth{-in} \psi_+\right) = 0 \ ,
\label{fermeq2}
\ee
\be
\eth{-i}\left(\dr -\frac{i}{r}\dth \pm n\frac{a}{r}\right)\psb_\pm +
  \eta f \eth{\mp in}\left(i\mu \psi_0 \pm \frac{g}{\sqrt{2}}\la \right) = 0 
\ .
\label{fermeq3}
\ee
The corresponding equations for the lower fermion components can be obtained 
from those for the upper components by complex conjugation, and 
putting $n \rightarrow -n$. The superconducting current corresponding to this 
solution (like \bref{witteq}, but with $\chi(t-z)$) is right moving. The 
angular dependence may be removed with the 
substitutions

\begin{eqnarray} 
\la & = & A(r)^\ast \eth{i(l-1)} \ , \\
\psi_+ & = & B(r)\eth{i(n-l)} \ , \\
\psi_- & = & C(r)\eth{-i(n+l)} \ , \\
\psi_0 & = & E(r)^\ast \eth{i(l-1)} \ .
\end{eqnarray}

For large $r$ the four solutions have the asymptotic forms

\bea
A(r) & \sim B(r)-C(r) &\sim e^{\pm gr} \ , \\
E(r) & \sim B(r)+C(r) &\sim e^{\pm \sqrt{2}\mu r} \ .
\eea
To be physically significant, solutions must be normalisable
\cite{Jackiw}, and so must be well behaved at $r=0$ and decay 
sufficiently rapidly as $r \rightarrow \infty$.
At small $r$ the least well behaved parts of the four solutions are  

\bea
A(r) & \sim & r^{l-1}  \ , \\
B(r) & \sim & r^{n-l}  \ , \\
C(r) & \sim & r^{-n-l} \ , \\
E(r) & \sim & r^{l-1} \ .
\eea
Thus in order to match up with
some combination of the two normalisable solutions at large $r$, at
least three of the small $r$ solutions must be well behaved at $r=0$. This
occurs when $1\leq l\leq |n|$, giving a total of $|n|$ independent solutions. 
Similarly, solutions for the lower components of the
fields also have $|n|$ independent solutions. In terms of the superconducting 
solution~(\ref{witteq}), these two sets of solutions correspond to currents 
flowing in opposite directions along the string. Note that the zero modes 
may also be enumerated using index theorems\cite{Jackiw}.

In general, in non-supersymmetric theories, it is difficult to find 
solutions for fermion zero modes in string backgrounds. However, in
the supersymmetric case, SUSY transformations relate the fermionic 
components of the superfields to the bosonic ones and we may use this to
obtain the fermion solutions in terms of the background string fields.
A SUSY transformation is 
implemented by the operator $G=e^{\xi Q + \xib \bar{Q}}$, 
where $\xi_{\alpha}$ are Grassmann parameters and $Q_{\alpha}$ are the 
generators of the SUSY algebra which we may represent by

\begin{eqnarray}
Q_\al & = & \frac{\partial}{\partial\th^\al} 
- i\sigma^\mu_{\al \ald} \thb^\ald \dmu \ , \\
\bar{Q}^\ald & = & \frac{\partial}{\partial\thb_\ald}
		- i\bar{\sigma}^{\mu \ald \al} \th_\al \dmu \ .
\label{SusyTransform}
\end{eqnarray}
In general such a transformation will induce a change of gauge. It is then 
necessary to perform an additional gauge transformation to return to the
Wess-Zumino gauge in order to easily interpret the solutions. For an abelian 
theory, supersymmetric gauge transformations are of the form

\begin{eqnarray}
\Phi_i & \rightarrow & e^{-i\Lambda q_i}\Phi_i \ , \\
\bar{\Phi}_i & \rightarrow & e^{i\bar{\Lambda} q_i}\bar{\Phi}_i \ , \\
V & \rightarrow & V + \frac{i}{g}\left(\Lambda - \bar{\Lambda}\right) \ ,
\end{eqnarray}
where $\Lambda$ is some chiral superfield. 

Consider performing an infinitesimal SUSY transformation on \bref{StringSol},
using $\dmu A^\mu = 0$. The appropriate $\Lambda$ to return to Wess-Zumino 
gauge is

\begin{equation}
\Lambda = ig\xib\sigb{\mu}\th A_\mu (y) 
\end{equation}
The component fields then transform in the following way

\begin{eqnarray} 
\phi_\pm(y) & \rightarrow & \phi_\pm(y)  
+ 2i\th\sm\xib D_\mu \phi_\pm(y) \ , \\ 
\th^2 F_0(y) & \rightarrow & \th^2 F_0(y) + 
2\th\xi F_0(y) \ , \\
-\th\sm\thb A_\mu(x) & \rightarrow &  
	-\th\sm\thb A_\mu(x) \nonumber \\
&& {}+ i\th^2 \thb\frac{1}{2} \sigb{\mu}\sigma^\nu \xib F_{\mu \nu}(x) 
- i\thb^2 \th\frac{1}{2} \sm\sigb{\nu} \xi F_{\mu \nu}(x)
\ .
\end{eqnarray}
Writing everything in terms of the background string fields,
only the fermion fields are affected to first order by the transformation.
These are given by

\begin{eqnarray}
\la_\al &\rightarrow& \frac{2na'}{gr}i(\sgz)^\beta_\al \xi_\beta \ , \\
(\psi_\pm)_\al &\rightarrow& \sqrt{2} \left(if'\sgr \mp
\frac{n}{r}(1-a)f \sgth\right)_{\al \ald} \xib^\ald \eta \eth{\pm in} \ , \\
(\psi_0)_\al &\rightarrow& \sqrt{2}\mu\eta^2(1-f^2)\xi_\al \ ,
\end{eqnarray}
where we have defined

\begin{eqnarray}
\sgth & = & \mat{0}{-i\eth{-i}}{i\eth{i}}{0} \ , \\
\sgr & = & \mat{0}{\eth{-i}}{\eth{i}}{0} \ .
\end{eqnarray}

Let us choose $\xi_{\alpha}$ so that only one component is nonzero.
Taking $\xi_2 = 0$ and $\xi_1 = -i\delta/(\sqrt{2}\eta)$, where $\delta$
is a complex constant, the fermions become

\begin{eqnarray}
\la_1 & = & \delta\frac{n\sqrt{2}}{g\eta}\frac{a'}{r} \ , \\ 
(\psi_+)_1 & = & \delta^\ast \left[f'+\frac{n}{r}(1-a)f\right]\eth{i(n-1)} 
\ , \\
(\psi_0)_1 & = & -i\delta\mu\eta(1-f^2) \ , \\
(\psi_-)_1 & = & \delta^\ast \left[f'-\frac{n}{r}(1-a)f\right]\eth{-i(n+1)} \ .
\end{eqnarray}
It is these fermion solutions which are responsible for the string
superconductivity.
Similar expressions can be found when $\xi_1 = 0$. It is clear from
these results that the string is not invariant under supersymmetry,
and therefore breaks it. However, since $f'(r), a'(r), 1-a(r)$ and $1-f^2(r)$ 
are all approximately zero outside of the string core, the SUSY breaking and
the zero modes are confined to the string. We note that this method gives us 
two zero mode solutions. Thus, for a winding number one string, we obtain the
full spectrum, whereas for strings of higher winding number, only a partial
spectrum is obtained.

\section{Theory D: Nonvanishing Fayet-Iliopoulos Term}

Now consider theory D in which there is just one primary charged chiral
superfield involved in the symmetry breaking and a non-zero Fayet-Iliopoulos
term. In order to avoid gauge anomalies, the model must contain other 
charged superfields. These are
coupled to the primary superfield through terms in the superpotential such that
the expectation values of the secondary chiral superfields are dynamically 
zero. The secondary superfields have no effect on SSB and are invariant under
SUSY transformations. Therefore, for the rest 
of this section we shall concentrate on the primary chiral 
superfield which mediates the gauge symmetry breaking. 

Choosing $\kappa = -\frac{1}{2}g\eta^2$, the theory is spontaneously
broken and there exists a string solution obtained from the ansatz

\begin{eqnarray} 
\phi & = & \eta e^{in\vp}f(r) \ , \\
A_\mu & = & -\frac{2}{g} n \frac{a(r)}{r}\delta_\mu^\vp \ , \\
D & = & \frac{1}{2}g\eta^2 (1-f^2) \ , \\
F & = & 0 \ .
\end{eqnarray}
The profile functions $f(r)$ and $a(r)$ then obey the first order equations

\begin{equation}
f' = n\frac{(1-a)}{r}f
\label{fEqnII}
\end{equation}

\begin{equation}
n\frac{a'}{r} = \frac{1}{4}g^2 \eta^2 (1-f^2)
\label{aEqnII}
\end{equation}

Now consider the fermionic sector of the theory and perform a SUSY
transformation, again using $\Lambda$ as the gauge function to return to
Wess-Zumino gauge. To first order this gives

\begin{eqnarray}
\la_\al &\rightarrow&  \frac{1}{2}g\eta^2(1-f^2)i(I+\sgz)_\al^\beta\xi_\beta\\
\psi_\al &\rightarrow& 
\sqrt{2}\frac{n}{r}(1-a)f(i\sgr-\sgth)_{\al \ald} \xib^\ald \eta \eth{in}  
\end{eqnarray}
If $\xi_1=0$ both these expressions are
zero. The same is true of all higher order terms, and so the string is
invariant under the corresponding transformation. For other $\xi$,
taking $\xi_1= -i\delta/\eta$ gives
\bea 
\la_1  &=& \delta g \eta(1-f^2) \\
\psi_1 &=& 2\sqrt{2} \delta^\ast \frac{n}{r}(1-a)f \eth{i(n-1)} 
\eea
Thus supersymmetry is only half broken inside the string. This is in
contrast to theory F which fully breaks supersymmetry in the string
core. The theories also differ in that theory D's zero
modes will only travel in one direction, while the zero modes of theory F 
(which has twice as many) travel in both directions. In both theories the
zero modes and SUSY breaking are confined to the string core.

\section{The Super-Bogomolnyi Limit}
In non-supersymmetric theories it is usually difficult to find
solutions for fermion zero modes on cosmic string backgrounds. In such
theories one can, however, often obtain solutions in the 
{\it Bogomolnyi limit} which, in our theory, corresponds to choosing

\be
2\mu^2 = g \ .
\label{BL}
\ee
In this limit, the energy of the vortex saturates a topological bound, there
are no static forces between vortices and the 
equations of motion for the string fields reduce to a pair of coupled first
order differential equations.
It is a useful check of the solutions obtained in the previous sections to
confirm that they reduce to those already known in the Bogomolnyi limit.

Imposing~(\ref{BL}), equations~(\ref{fEqn},\ref{aEqn}) become

\begin{eqnarray}
f' & = & n \frac{f}{r}(1-a) \ , \\
n\frac{a'}{r} & = & \mu^2\eta^2(1-f^2) \ .
\label{Bogomolnyi}
\end{eqnarray}
Note that these are identical to~(\ref{fEqnII},\ref{aEqnII}) and that 
therefore all solutions to theory D are automatically Bogomolnyi solutions.
Imposing \bref{BL} on 
(\ref{fermeq1},\ref{fermeq2},\ref{fermeq3}) gives the following solutions.

\begin{eqnarray}
\la_1 & = & \delta\mu\eta(1-f^2) \ , \\
(\psi_+)_1 & = & 2\delta^\ast n\frac{f}{r}(1-a)\eth{i(n-1)} \ , \\
(\psi_0)_1 & = & -i\delta\mu\eta(1-f^2) \ , \\
(\psi_-)_1 & = & 0 \ .
\end{eqnarray}
This limit, with $n=1$, was considered for a similar theory by Garriga
and Vachaspati\cite{Garriga} and the above results are in agreement
with theirs. This is a useful check of the techniques we use.

\section{Concluding Remarks}

We have investigated the structure of cosmic string solutions to
supersymmetric abelian Higgs models. For completeness we have analysed two 
models, differing
by their method of spontaneous symmetry breaking. However, we expect theory F 
to be more representative of general
defect forming theories, since the SSB employed there is not specific to 
abelian gauge groups. 

We have shown that although SUSY remains unbroken outside the
string, it is broken in the string core (in contrast to the gauge
symmetry which is restored there). In theory F supersymmetry is broken 
completely in the string core by
a nonzero $F$-term, while in theory D supersymmetry is partially
broken by a nonzero $D$-term. We have demonstrated that, due to the
particle content and couplings dictated by SUSY, the cosmic
string solutions to both theories are superconducting in the Witten sense. 
Thus, all supersymmetric abelian cosmic strings are superconducting due to 
fermion zero modes. 

Although explicitly solving for such zero modes is difficult in the
case of non-supersymmetric theories, in the models we study it is possible
to use SUSY transformations to relate the functional form of
the fermionic solutions to those of the background string fields, which are
well-studied. For theory D the solutions all obey the Bogomolnyi equations 
exactly, and for theory F we have also checked that the solutions we find 
reduce to those already known in the special case of the Bogomolnyi limit.

While we have performed this first analysis for the toy model of an abelian 
string, we expect the techniques to be quite general and in fact to be more
useful in non-abelian theories for which the equations for the fermion zero
modes are significantly more complicated. The question of superconductivity
in non-abelian SUSY cosmic strings is under investigation.

There remain many unanswered questions concerning supersymmetric topological 
defects and the cosmological implications of particle physics theories
which admit them. While this work was in preparation, two 
papers\cite{{D&S},{H&K}} appeared in which SUSY topological defects
were considered in different settings to our work. 
We are currently investigating other roles that
supersymmetric topological defects may play in the early universe. Clearly, 
there is much scope for further study.

\acknowledgments
We would like to thank Sean Carroll, Dan Freedman, Ruth Gregory, Markus Luty, 
Hugh Osborn, Malcolm Perry, Patrick Peter, Tanmay Vachaspati and Mark Wise 
for helpful discussions.

A.C.D. and M.T. would like to thank the Aspen Center for Physics, where some
of this work was done, for support and hospitality.

This work is supported in part by PPARC and the
E.U. under the HCM program (CHRX-CT94-0423) (S.C.D. and A.C.D.), Trinity
College Cambridge (S.C.D.) and by funds provided by 
the U.S. Department of Energy (D.O.E.) under cooperative research agreement
\# DF-FC02-94ER40818 (M.T.).


\begin{thebibliography}{100}
\bibitem{structure}
A. Vilenkin, {\it Phys. Rev. Lett.} {\bf 46}, 1169 (1981); \\
R. Brandenberger, L. Perivolaropoulos and A. Stebbins, 
{\it Int. J. Mod. Phys.} {\bf A}, Vol. 5, No. 9, 1633 (1990); \\
T. Vachaspati and A. Vilenkin, {\it Phys. Rev. Lett.} {\bf 67}, 1057 (1991); \\
D. Vollick, {Phys. Rev.} {\bf D45}, 1884 (1992).

\bibitem{DMEWBG}
R. Brandenberger, A.-C. Davis and M. Hindmarsh, {\it Phys. Lett.} {\bf B263}, 
239 (1991); \\
R. Brandenberger and A.-C. Davis, {\it Phys. Lett.} {\bf B308}, 79 (1993); \\
R. Brandenberger, A.-C. Davis, and M. Trodden, {\it Phys. Lett.} {\bf B335},
123 (1994); \\
R. Brandenberger, A.- C. Davis, T. Prokopec  and M. Trodden, 
{\it Phys. Rev.} {\bf D53}, 4257 (1996); \\
T. Prokopec, R. Brandenberger, A.-C. Davis and M. Trodden,
{\it Phys. Lett.} {\bf B384}, 175 (1996); \\
M. Trodden, A.-C. Davis and R. Brandenberger, {\it Phys. Lett.} {\bf B349}, 
131 (1995).

\bibitem{PQ}
R.D. Peccei and H.R. Quinn, {\it Phys. Rev. Lett.} {\bf 38}, 1440 (1977).

\bibitem{B&S}
R.A. Battye and E.P.S. Shellard, {\it Phys.Rev. Lett.} {\bf 73}, 2954 (1994); 
\\
{\it ibid} {\bf 76}, 2203 (1996).

\bibitem{Witten}
E. Witten, {\it Nucl. Phys.} {\bf B249}, 557 (1985).

\bibitem{vortons}
R. Brandenberger, B. Carter, A.-C. Davis and M. Trodden, {\it Phys. Rev.}
{\bf D54}, 6059 (1996).

\bibitem{Kibble}
T.W.B. Kibble, {\it J. Phys.} {\bf A9}, 1387 (1976).

\bibitem{Morris} 
J.R. Morris, {\it Phys. Rev.} {\bf D53}, 2078 (1996).

\bibitem{Garriga} 
J. Garriga and T. Vachaspati, {\it Nucl. Phys.} {\bf B438}, 161 (1995).

\bibitem{Fayet I}
P. Fayet, {\it Nuovo Cimento} {\bf A31}, 626 (1976).

\bibitem{susyBook} 
J. Wess and J. Bagger, {\em Supersymmetry and Supergravity},
second edition (Princeton University Press, Princeton, NJ, 1992).

\bibitem{Fayet II} 
P. Fayet and J. Iliopoulos, {\it Phys. Lett.} {\bf 51B}, 461 (1974).

\bibitem{NO}
H. Nielsen and P. Olesen, {\it Nucl. Phys.} {\bf B61}, 45 (1973).

\bibitem{Jackiw} 
R. Jackiw and P. Rossi, {\it Nucl. Phys.} {\bf B190}, 681 (1981).

\bibitem{D&S}
G. Dvali and M. Shifman, {\it Dynamical Compactification as a Mechanism of
Spontaneous Supersymmetry Breaking}, hep-th/9611213, November 1996.

\bibitem{H&K}
R. Holman and S. Prem Kumar, ``{\it Gravitino Zero Modes on $U(1)_R$ 
Strings}'', hep-ph/9702011, February 1997.

\end{thebibliography}
\end{document}